\documentclass[prx,preprint]{revtex4-2}
\usepackage{graphicx} 
\usepackage{amsmath}
\usepackage{wrapfig}
\usepackage[sort&compress]{natbib}
\usepackage[T1]{fontenc}
\usepackage{color}
\usepackage{lmodern}        
\DeclareUnicodeCharacter{0308}{\"{}}
\usepackage{enumitem}
\usepackage{xcolor}
\usepackage[utf8]{inputenc}
\usepackage{hyperref}
\usepackage{float}
\usepackage{hyperref}
\hypersetup{colorlinks,allcolors=blue}

\begin{document}
\newcommand{\comm}[1]{{\color{red}{#1}}}
\newcommand{\arghacomm}[1]{{\color{green}{#1}}}
\newcommand{\sgvcomm}[1]{{\color{violet}{#1}}}

\title{Self-diffusion is temperature independent on active membranes}
\author{Saurav G. Varma}
\email{sauravvarma@iisc.ac.in}
\author{Argha Mitra}
\email{arghamitra@iisc.ac.in}
\author{Sumantra Sarkar}
\email{sumantra@iisc.ac.in}
\affiliation{Center for Condensed Matter Theory, Department of Physics, Indian Institute of Science, Bengaluru, Karnataka, 560012}
\date{\today}
\begin{abstract}
 Molecular transport maintains cellular structures and functions. For example, lipid and protein diffusion sculpts the dynamic shapes and structures on the cell membrane that perform essential cellular functions, such as cell signaling. Temperature variations in thermal equilibrium rapidly change molecular transport properties. The coefficient of lipid self-diffusion increases exponentially with temperature in thermal equilibrium, for example. Hence, in the noisy cellular environment, where temperatures can fluctuate widely due to local heat generation, maintaining cellular homeostasis through molecular transport is hard in thermal equilibrium. In this paper, using both molecular and lattice-based modeling of membrane transport, we show that the presence of active transport originating from the cell's cytoskeleton can make the self-diffusion of the molecules on the membrane robust to temperature fluctuations. The resultant temperature-independence of self-diffusion keeps the precision of cellular signaling invariant over a broad range of ambient temperatures, allowing cells to make robust decisions. We have also found that the Kawasaki algorithm, the widely used model of lipid transport on lattices, predicts incorrect temperature dependence of lipid self-diffusion in equilibrium. We propose a new algorithm that correctly captures the equilibrium properties of lipid self-diffusion and reproduces experimental observations.

\end{abstract}
\maketitle
\section{Introduction}
Molecular transport is essential for various physical and biological processes. In cells, where molecules are the primary drivers of all structure and function, spatial organization and transport of molecules determine the robustness and accuracy of various cellular processes. For example, diffusion, one of the primary modes of transport for cellular molecules, imposes a fundamental limit on the cell's ability to sense its environment \cite{berg1977physics}. Hence, understanding the factors that govern the molecular transport processes in cells, such as molecular crowding \cite{javanainen2017diffusion, fabian2023protein} or interaction with the actin cytoskeleton (ACS)~\cite{kusumi2005paradigm, ritchie2005detection, eggeling2009direct, mueller2011sted}, is of general interest. In this paper, we investigate the effect of the ACS on lipid transport on the plasma membrane and show that the nonequilibrium (active) forces imparted by ACS on the lipids fundamentally change the nature of molecular transport on the plasma membrane (PM).

Cellular transport primarily happens through two mechanisms: diffusion, which obeys the detailed balance principle, and active transport, which breaks detailed balance \cite{bowick2022symmetry}. The constraint of detailed balance is a defining feature of an equilibrium system. It states that the probability flux from a state $i$ to a state $j$ exactly equals the flux from state $j$ to state $i$. Mathematically, 
\begin{eqnarray}
    p_i \pi_{ij} &=& p_j \pi_{ji},
\end{eqnarray}
where $p_{i(j)}$ is the occupation probability of state $i (j)$ and $\pi_{ij}$ is the transition probability from state $i$ to state $j$. In cells, active transport is facilitated by motor proteins, which consume ATP to drive various cellular structures out of equilibrium. For example, they interact with the cortical ACS, a meshwork of actin filaments adjoining the cell's PM, and make them active. The interaction of the active ACS with the PM leads to patterns, structures, and behaviors that are not found in thermal equilibrium \cite{ritchie2003fence, oda2013dynamic, koster2016cortical}. These changes are brought in by the active transport of various molecules on the plasma membrane (PM) \cite{beemiller2013regulation}, whose repercussions are often significantly different from its passive counterpart \cite{goswami2008nanoclusters}. 

As a concrete example, let's consider the diffusion coefficient's temperature ($T$)-dependence. Diffusion of a free colloid obeys the Stokes-Einstein relationship, which states that $D(T) = k_BT/\gamma$, where $k_B$ is the Boltzmann constant and $\gamma$ is the friction coefficient. Also, the diffusion of molecules adsorbed on periodic surfaces is determined by escape from the periodic potential generated by the surface molecules, which results in an Arrhenius-like form for the diffusion coefficient: 
\begin{eqnarray}\label{eq:2}
    D(T) &=& D_0 e^{-\frac{E_A}{k_BT}},
\end{eqnarray}
where $E_A$ is proportional to the surface potential \cite{montalenti1999jumps}. Self-diffusion of molecules also follows the Arrhenius kinetics, where the many-body interaction between the molecules determines the $E_A$. Together, we will call them activated diffusion. Activated diffusion is often observed on model membranes, where $E_A$ is of the order of tens of $k_BT$ \cite{javanainen2010free, debnath2013simulation, bag2014temperature}. In contrast, the $T$-dependence of diffusion is unclear on live cell plasma membrane, where active effects are predominant. 

Recently, it was observed that, in live cells, Glycosylphosphatidylinositol anchored proteins (GPI-AP), a model cell surface protein and a known interaction partner of the ACS~\cite{goswami2008nanoclusters}, shows $T$-independent diffusion \cite{saha2015diffusion}. This observation contradicts prior works, which show, using very similar techniques (Fluorescence Correlation Spectroscopy: FCS), that diffusion is $T$-dependent even in live cells~\cite{machavn2010lipid,favard2011fcs, bag2014temperature, lee2015live}. Therefore, it remains unclear whether the transport of molecules in live cells is $T$-independent or not. An important observation is that the $T$-independence of diffusion is dependent on the scale of observation~\cite{saha2015diffusion}. When the membrane is observed with a small FCS spot ($3\times 10^4 ~ nm^2$), diffusion is $T$-independent. However, as the spot size increases ($6\times 10^4 ~ nm^2$), diffusion becomes $T$-dependent \cite{saha2015diffusion}. Therefore, we hypothesize that processes that are important at the nanometer scale, such as the interaction of the membrane with the ACS, are the key drivers of $T$-independent transport of molecules. In this paper, we test this hypothesis using a coarse-grained molecular dynamic (CG-MD) model and a lattice model of model membranes. 

Now, we briefly describe the results and the conclusions of the paper. The paper is broadly divided into two parts. In the first part, we ran equilibrium CG-MD simulations to study lipid self-diffusion in model membranes, which showed that diffusion is an activated process irrespective of the observation scale. In the second part of the paper, we tested the effect of an underlying nonequilibrium driving on the activated diffusion process. Specifically, we studied the effect of ACS on the lipid self-diffusion using a lattice model of lipids. We find that coupling to the ACS makes lipid diffusion $T$-independent. Surprisingly, lipids that do not couple to the ACS can also show near $T$-independence because of their interactions with the ACS-coupled lipids. We comment on the consequences of the $T$-independence and its biological relevance to end the paper. We have also found that the Kawasaki algorithm, which is used extensively to simulate lipid diffusion, does not accurately capture diffusion's thermodynamics. We propose an alternate algorithm, called the barrier hop dynamics, which accurately captures the equilibrium kinetics of activated diffusion.

\section{Results}
\subsection{Bulk diffusivity in equilibrium is $T$-dependent} 

\begin{figure}[ht]
    \centering
    \includegraphics[width=\columnwidth]{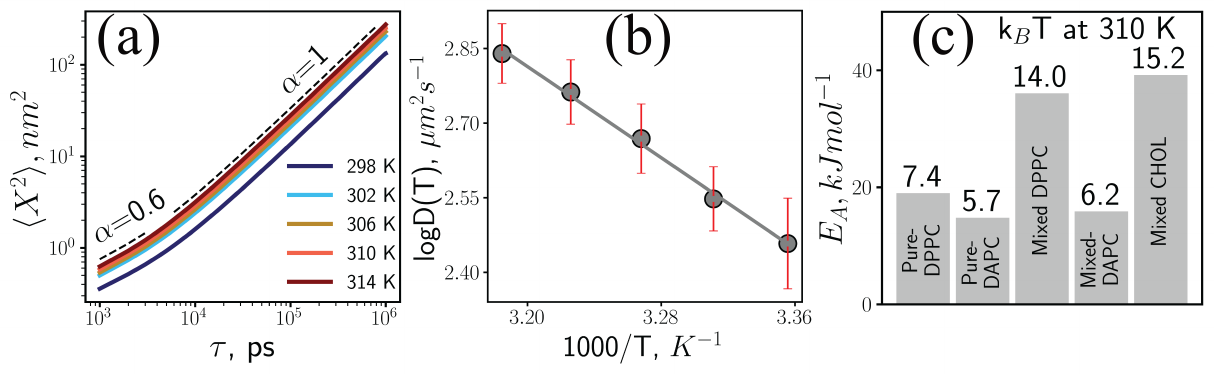}
    \caption{Activated diffusion in CG-MD model.  (a-b) MSD ($\langle X^2 \rangle$) and diffusivity (semilog scale) of pure DPPC membrane at different $T$ (298 K - 314 K). (c) The activation energy ($E_A$) of lipids for different PM models  $k_B$T at 310 K. All the results from long time scale simulations. Error bars are the standard deviation of 8 replicates. In (b) $\log D(T)$ is plotted vs $1000/T$, as per experimental convention~\cite{bag2014temperature}.}
    \label{fig:diffusive_arrhenius}
\end{figure}

We first checked the behavior of the bulk transport coefficients of lipids as a function of temperature, $T$, using CG-MD simulation (see section \ref{CG-MD}). We computed the lipids' time and ensemble-averaged mean squared displacement (MSD) to measure the bulk transport coefficients (see \ref{MSD_cal}). The MSD curves showed clear $T$-dependence (Fig.~\ref{fig:diffusive_arrhenius}a, S1, S2). Also, all MSD curves had the characteristic subdiffusive regions for short lag time, $\tau$, with an anomalous coefficient of $0.6$ \cite{kneller2011communication}, and at long times, showed diffusive transport. We measured the diffusion coefficients at different temperatures, which varied according to the Arrhenius form (Eq:~\ref{eq:4}, \ref{eq:5} and Fig.~\ref{fig:diffusive_arrhenius}b), as shown in various experiments before~\cite{macedo1965relative, vaz1985translational, almeida1992lateral, filippov2003effect}. The activation energy, $E_A$ ranges between $6-14~k_BT$ (Fig~\ref{fig:diffusive_arrhenius}c), which is in the same ballpark of experimental~\cite{bag2014temperature} and atomistic simulation data reported earlier~\cite{apajalahti2010concerted, javanainen2010free}, which provides an independent check on the correctness of our simulations. Therefore, we conclude that, in equilibrium, lipid self-diffusion is an activated process and depends strongly on temperature.

\subsection{Local diffusivity in equilibrium is $T$-dependent}
\begin{figure}[ht]
    \centering
    \includegraphics[width=\columnwidth]{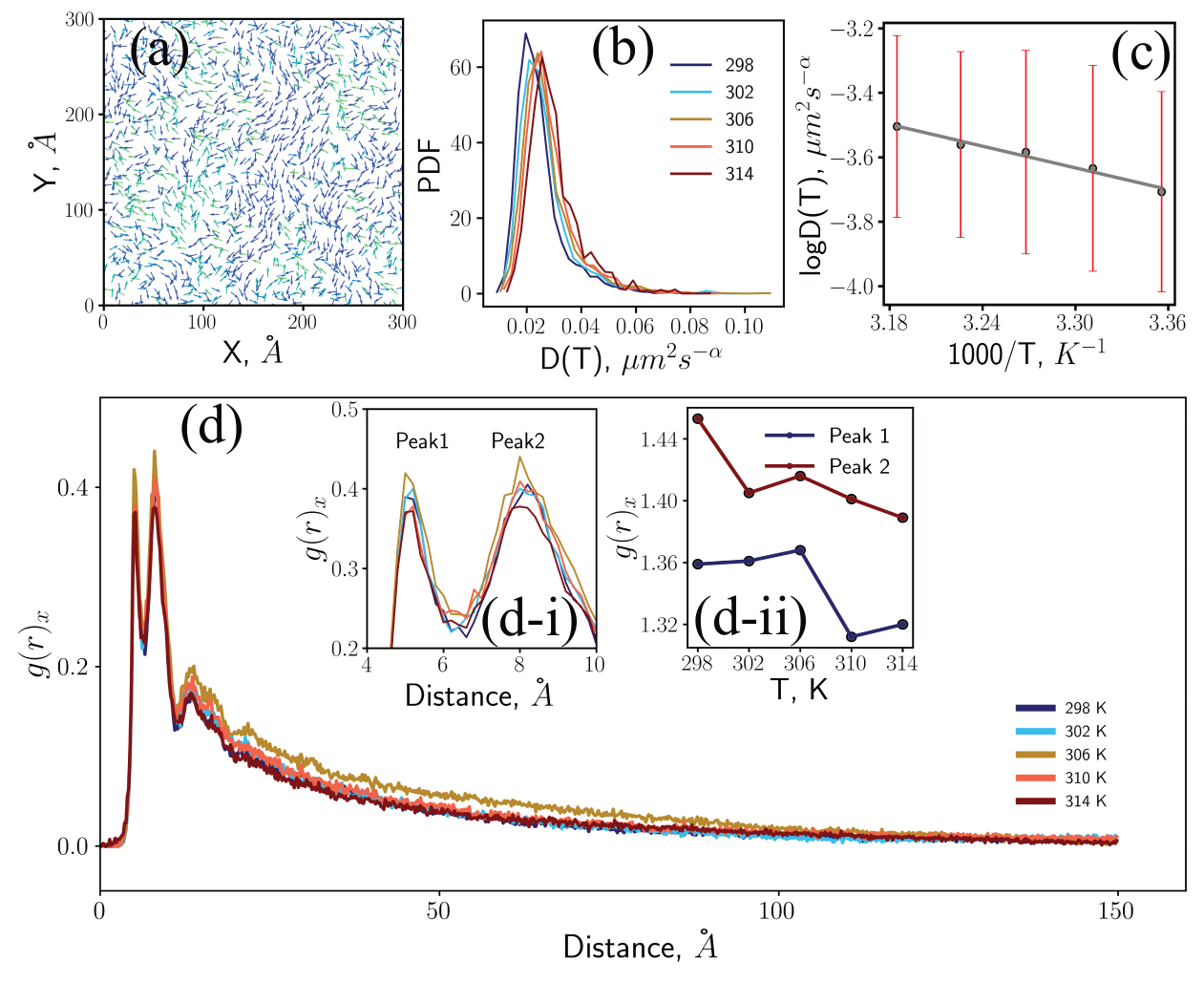}
    \caption{Nanoscale transport of lipids is $T$-dependent. (a) Snapshot of lateral displacement of DPPC lipids at 314 K with $\tau$ = 20 ps. (b) Probability distribution of diffusivity in the grid-box ($3~nm \times~3~nm$) at different $T$, ($\alpha$ = 0.55). (c) Transport coefficient, D($T$) at different T in semilog scale, ($\alpha$ = 0.55). (d) Displacement-displacement correlation (DDC) quantification for DPPC to its respective T. (d-i)  Zoomed view of Peak 1 and Peak 2 from DDC. (d-ii) Peak height at Peak 1 and 2 with its corresponding $T$. All the results are from short-time scale simulations. Error bars are standard deviation of 20 replicates.}
    \label{fig:Correlated_flow}
\end{figure}

In ref.~\cite{saha2015diffusion}, $T$-independent transport arose only when the spot area of the FCS measurement was small (confocal spot area $\omega^2 = 3 \times 10^4 ~ nm^2$). Our model membrane is smaller than this threshold, so if temperature independence were only a matter of length scale, our simulations would have captured it. However, because MSD is averaged over many ensembles and is a bulk measure, it may hide local $T$-independent transport processes. Hence, we investigated the transport of lipids at various length scales, ranging from individual lipids to regions containing 10-15 lipids. 
 
We found that individual lipids showed correlated motion with their neighbors and next-nearest neighbors when the lag time was short (ps-ns) (Fig~\ref{fig:Correlated_flow}a), which has also been shown before \cite{falck2008lateral}. In contrast, in the long time scales where we examined ensemble-averaged MSD curves (Fig.~\ref{fig:diffusive_arrhenius}a), we noticed diffusive behavior implying random movement of lipids. Therefore, the bulk, long-time behavior of the lipid transport, as seen in the MSD curves, is strikingly different than their local short-time transport. Hence, we analyzed the local nanoscale transport through means different from ensemble-averaged MSD. 

First, we measured the local diffusivities of the lipids using their MSD averaged over grids of size 3 nm $\times$ 3 nm (Fig. S3). From these distributions, we computed the distribution of the transport coefficient, $D(T)$, at different temperatures and measured their average value. Both showed weak but clear Arrhenius-like $T$-dependence (Fig.~\ref{fig:Correlated_flow}b,~c). Therefore, in equilibrium, even at the nanoscale, we observed clear signatures of activated diffusion, which consolidated our hypothesis that active driving is needed to make activated diffusion $T$-independent. 

Because MSD based measurements of transport coefficients are defined in the $t\rightarrow \infty$ limit, it is unclear whether MSD-based measurement of $D(T)$ is accurate or not. A more accurate alternative is to measure the $T$-dependence through the correlation functions (see \ref{correlation_cal}) of the lipid transport. For this purpose, we measured the displacement-displacement correlation function \cite{poole1998spatial, bennemann1999growing, donati1999growing, ediger2000spatially, richert2002heterogeneous, narumi2008simulation, narumi2009study, puosi2012spatial}, which showed two prominent peaks at around 0.51 and 0.83 nm distance (Fig.~\ref{fig:Correlated_flow}d). The peaks showed weak but systematic variation with temperature: both decreased in height with increasing temperature (Fig.~\ref{fig:Correlated_flow}d-i, d-ii). However, the variation is so small that it will be impossible to detect it in present experiments. 

Our observations suggest that equilibrium correlated flows cannot render the transport $T$-independent. Hence, we conjectured that nonequilibrium flows of lipids must be the origin of $T$-independence in the FCS experiment. At the smallest lengthscales probed by the FCS experiments \cite{pyenta2001cross, ritchie2005detection}, this nonequilibrium flow must affect tens of thousands of lipids simultaneously, indicating that the cortical ACS is a possible driver of such flows \cite{goswami2008nanoclusters, gowrishankar2012active, saha2022active}. Unfortunately, it is tough to investigate the coupled dynamics of the ACS with the membrane using CG-MD because of the prohibitive computational cost. To the best of our knowledge, even with the best supercomputers, the interaction between membrane and only a single actin filament has been studied \cite{schroer2020charge}. Hence, we resort to a mesoscopic lattice model \cite{das2016phase} of the membrane to investigate the effect of the ACS on the membrane-lipid transport (See \ref{lattice model}).  

\subsection{Kawasaki algorithm gives incorrect equilibrium kinetics}

\begin{figure}[ht]
    \centering
    \includegraphics[width=\columnwidth]{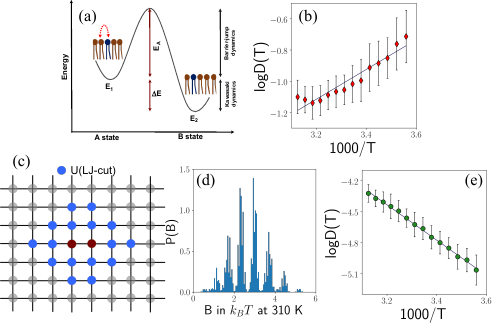}
    \caption{(a) The energy profile diagram for the exchange process of the lipids from, let's say, A state to B state, where $\Delta E $ and $E_A$ are the energies used in the Boltzmann factor for the Kawasaki (KD) and Barrier Hop Dynamics (BHD), respectively. (b) Kawasaki dynamics predicts unphysical variation of $D(T)$ with $T$. (c) Illustration of the lattice model, where maroon colored lipids at the center are undergoing exchange move and interacting with lipids in blue through truncated LJ potential. (d) The probability distribution of barrier heights ($B=-U$) in $k_BT$ at 310 K  resulting from various configurations uniformly sampled (e) BHD gives us the desired activated diffusion given by Arrhenius kinetics \cite{bag2014temperature} $E_A = 5.81~k_BT$ at 310K.}
    \label{fig:KD_vs_BHD}
\end{figure}

Traditionally, the local Kawasaki algorithm is used to model the self-diffusion of lipids on lattice models \cite{kawasaki1972phase, machta2011minimal, kimchi2018ion, das2016phase}. However, the Kawasaki algorithm is not appropriate for modeling self-diffusion because it cares only about the energy difference between the state before and after the swap and does not care about the energies of the transition states that facilitate the move (Fig.~\ref{fig:KD_vs_BHD}a). Hence, it correctly captures thermodynamics but fails to capture the kinetics of any processes. Here, we show this discrepancy for the self-diffusion of lipids in a membrane.

In the membrane, a lipid interacts with its neighboring lipids. When two adjacent lipids swap places, there are significant rearrangements of the local lipid configurations. Such \textit{transition states} usually have higher free energy than the configurations before and after the swap. From a modeling perspective, it implies that every lipid swap requires overcoming a kinetic barrier with activation energies correlated with the diffusing lipid's total interaction energy with its neighbors. In Kawasaki algorithm, we ignore this activation energy. The resultant transport process is unphysical: the coefficient of self-diffusion shows nonmonotonic variations with $T$ (Fig.~\ref{fig:KD_vs_BHD}b). For temperatures higher than 312 K, diffusivity decreases with $1/T$, and below 312 K, diffusivity increases with $1/T$, violating experimental and computational observations \cite{edidin1977effect, bag2014temperature, wang2016dppc}, where diffusivity decreases monotonously with $1/T$. 

Kawasaki dynamics gives such inaccurate results because it incorrectly increases the swap rate of like lipids in lipid domains, where swapping does not cost any energy. However, because like lipids attract each other, the energetic cost of kinetic rearrangements is high, resulting in a high activation barrier. Therefore, kinetically, we expect like lipids to slow down in a lipid domain, as seen in experiments \cite{ritchie2005detection}. However, Kawasaki algorithm is agnostic of this cost and leads to inaccurate $T$-dependence at low temperatures (high $1/T$), where lipid domains form easily. At higher temperatures (low $1/T$), lipids of various species are well-mixed, and the kinetic barriers are close to zero. Hence, there, the Kawasaki algorithm predicts the right temperature dependence.

\subsection{Barrier Hop Dynamics (BHD) gives correct equilibrium kinetics}

To incorporate the kinetic effects into our model, we use a simplified model of activation barrier, where the barrier height is equal to the negative of the total interaction energy of the diffusing lipid as shown in Fig. \ref{fig:KD_vs_BHD}a. 
\begin{eqnarray}
    B &=& -U 
\end{eqnarray}
We call this algorithm Barrier Hopping Dynamics (BHD). Similar algorithms have been used before to model activated processes~\cite{landau2021guide}; they provide excellent qualitative insights. Incorporating more details about the transition state and the associated free energies in the model may improve the accuracy of the model, but it does not guarantee better quantitative accuracy of the results. Because the lattice model is itself an ultra-coarse-grained description of the membrane lipids, it is unclear whether it is even possible to identify correct transition states or, if possible, whether they provide an accurate description of the microscopic activation processes.

 Using the BHD algorithm (see \ref{BHD_moves}), we model the equilibrium self-diffusion of lipids on the lattice. The lipids interact with each other through Lennard-Jones potential, from which we calculate the total interaction energy, $U$, of a pair of lipids (Fig.~\ref{fig:KD_vs_BHD}c). $U$ depends on the local lipid configurations and there are $2^{18} \approx 2.6\times 10^5$ of them. From a random sampling of these configurations, we obtain the distribution of barrier heights, which vary between 0 and 6 $k_BT$ at 310 K (Fig.~\ref{fig:KD_vs_BHD}d). The self-diffusion coefficient obtained from this model follows Arrhenius kinetics with an activation barrier of $E_A = 5.81 k_BT$ at 310K (Fig.\ref{fig:KD_vs_BHD}e), which is within a factor of three of the mixed-membrane $E_A$ and comparable to single component membrane $E_A$ obtained from CG-MD simulations (Fig.~\ref{fig:diffusive_arrhenius}c). Therefore, in the lattice model, we have used the BHD algorithm to model all equilibrium diffusive moves.

\subsection{Active transport of lipids is $T$-independent}
 Through its interaction with motor proteins, ACS forms dynamic spatiotemporal patterns called asters, which are roughly circular in shape~\cite{gowrishankar2012active, das2016phase, mondal2023coarsening}. Because of the exchange of filamentous actin between the ACS and the cytosol, called actin turnover, the asters spontaneously form and decay with a characteristic lifetime, $\tau_a$~\cite{fritzsche2017self, sarkar2023lifetime}. Here, we use a simplified model of aster dynamics, described in \cite{das2016phase}, to model its effect on the membrane. We assume that all asters are circular and of the same size $R_a$. The lifetime of an individual aster is an exponential random variable with mean $\tau_a$. We conserve the number of asters and assume that aster creation and annihilation are uncorrelated events. Hence, as soon as an aster disappears, another aster forms at a random location~\cite{das2016phase}. To ensure the uncorrelated formation of asters, we use a sufficiently small density of asters, which does not affect our main conclusion. The effect of the asters on the membrane is captured by introducing patches where active moves are allowed. These patches have the same size and lifetime as the aster that adjoins it. Passive lipids bind reversibly to the ACS when they are within this patch. When bound, they move persistently toward the aster core via active MC moves. When not bound, passive lipids move diffusively through BHD. Inert lipids never bind to the ACS and always move diffusively through BHD. Outside this patch, all lipids move diffusively through BHD (see~\ref{active_moves})

\begin{figure}[ht]
    \centering
    \includegraphics[width=\linewidth]{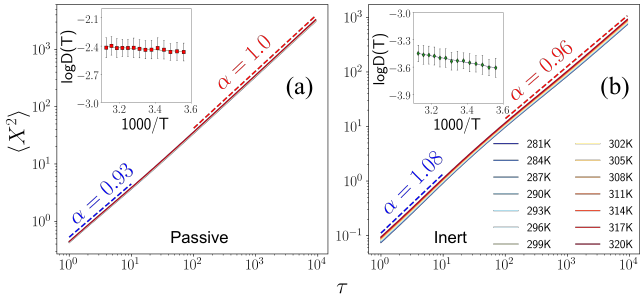}
    \caption{MSD, $\langle X^2 \rangle$, vs lagtime $\tau$ for passive and inert lipids for $p_b = 1$. (a) Passive lipids are subdiffusive at timescales comparable to average aster remodeling time and become diffusive at timescales longer than that. Inset: $\log D(T)$ vs $1000/T$ shows that the transport coefficient is nearly $T$-independent for passive lipids. (b) Inert lipids, in contrast, show a transition from weakly superdiffusive to weakly subdiffusive transport at the same timescale. Inset: For inert lipids, $D(T)$ increases exponentially with $T$. The $T$-(in)dependence is also evident from the spread of the MSD curves with $T$.}
    \label{fig:p_b_1}
\end{figure}

The appearance and disappearance of the asters introduce active fluctuations in the lipid movement, which changes their transport properties. To illustrate this change, we studied an extreme system where the passive lipids, once inside an aster patch, remained always bound to the aster and moved only through active moves. We found that the MSD of the passive lipids showed negligible $T$-dependence (Fig.~\ref{fig:p_b_1}a), implying the $T$-independence of $D(T)$ (Fig.~\ref{fig:p_b_1}a-inset). Therefore, our hypothesis that active fluctuations lead to T-independent diffusivity seems correct.

To check whether such weak variation can be captured in experiments, we computed the $Q_{10}$ metric \cite{wey1981lateral, saha2015diffusion}. To have significant $T$-dependence $Q_{10}$ must be greater than $1.2$ \cite{saha2015diffusion}. However, we found that $Q_{10}$  for the passive lipids was lower than this threshold. Hence, its $T$-dependence will not be detected in experiments.   

The inert lipids, which never couple to the ACS, also show altered transport properties, with weak superdiffusive behavior at small $\tau$ and weak subdiffusive behavior at large $\tau$ as shown in Fig. \ref{fig:p_b_1}b. Therefore, to investigate the $T$-dependence of the transport, we cannot use diffusivity and must use a general transport coefficient. In general, we can write: 
\begin{equation}
	\langle X^2\rangle = 4D(T)\tau^{\alpha},
\end{equation}
where $\langle X^2\rangle$ is the MSD, $\alpha$ is the anomalous diffusion exponent, and $D(T)$ is a transport coefficient equal to the diffusivity when $\alpha = 1$. We use this general definition of the transport coefficient, $D(T)$, to quantify the $T$-dependence of the lipids. As shown in Fig.~\ref{fig:p_b_1}b-inset, $D(T)$ for inert lipids shows a much stronger dependence on temperature than the passive lipids. This observation consolidates our hypothesis that ACS activity is the driver of $T$-independent transport, and it also shows that the transport of a lipid can be affected even when it does not interact with the ACS directly. 

\begin{figure}[h!]
    \centering\includegraphics[width=\columnwidth]{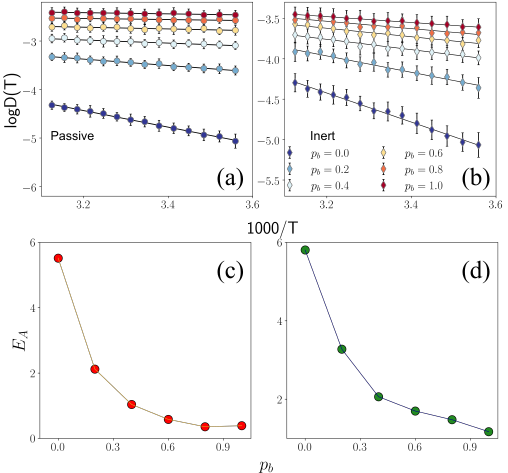}
    \caption{Effect of finite binding affinity between the passive lipids and the ACS. (a) $D(T)$ for passive lipids for different $p_b$. $T$-dependence is recovered at lower $p_b$. (b) $D(T)$ for inert lipids shows stronger variation with $T$. (c) $E_A$ for passive lipids, and (d) $E_A$ for inert lipids. $E_A$ decreases with increasing $p_b$, that is transport becomes more $T$-independent with increasing $p_b$.}
    \label{fig:aff_var}
\end{figure}

To test our hypothesis further, we systematically reduced the binding probability, $p_{b}$ of the passive lipids with the ACS. We found that for low $p_b$, the equilibrium $T$-dependence was recovered, with the exponential variation of $D(T)$ with temperature for both passive and inert lipids (Fig.~\ref{fig:aff_var}). At high $p_b$, $D(T)$ became $T$-independent for passive lipids and weakly $T$-dependent for inert lipids. These observations provided the final confirmation that active, nonequilibrium, fluctuations are responsible for $T$-independent transport in our system.

\subsection{Transport of inert lipids}
\begin{figure}[ht]
    \centering
    \includegraphics[width=\columnwidth]{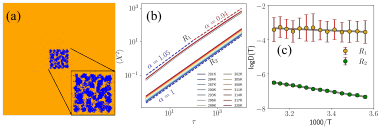}
    \caption{Transport of inert lipids in a composite region. (a) The snapshot shows the entire simulation box $500\times500$ grids, and the region in the middle (and the zoomed-in inset) is $R_1$ ($100\times100$ grids). Regions outside $R_1$ is $R_2$. Inert lipids are shown in orange, and passive lipids are shown in blue. $R_1$ contains both passive and inert lipids and $R_2$ contains only inert lipids. (b) MSD, $\langle X^2 \rangle$, vs $\tau$ (lag time) and (c) $\log D(T)$ vs $1000/T$ for  $T \in [281 K, 321 K]$ for inert lipids in the two regions.}
    \label{fig:aster_patch_analysis}
\end{figure}

As shown in Fig. \ref{fig:p_b_1}, the transport of inert lipids is superdiffusive at short $\tau$, even though there is no direct coupling to the active movements. We hypothesized that superdiffusion arises indirectly due to the excluded volume interaction between inert and passive lipids and the directed movement of the passive lipids in an aster. Before an aster forms, both types of lipids are equally likely to be present in a given region. However, when an aster forms, the passive lipids are more likely to cluster around the aster cores because of their directed movement towards the core and the expulsion of the inert lipids from the core due to excluded volume interaction. To test this idea, we created a system where asters form only in a small region at the center of the lattice ($R_1$). Both passive and inert lipids populate $R_1$, whereas only inert lipids populate the area outside this subsystem ($R_2$) as shown in Fig. \ref{fig:aster_patch_analysis}a. After a sufficiently long time, we observed distinct transport behavior of inert lipids in regions $R_1$ and $R_2$. Because $R_1$ resembles the lattice model we have used so far, we expect the MSD of the inert lipids in $R_1$ to be similar to Fig. \ref{fig:p_b_1}b. Whereas in  $R_2$, they only diffuse with the BHD moves, which should make the transport the same as in an equilibrium system, with strong $T$-dependent activated diffusion. We indeed observe the same dependence for MSD (Fig. \ref{fig:aster_patch_analysis}b). Moreover, the $D(T)$ looks strikingly different in the two regions: it is almost $T$-independent in $R_1$ but shows strong $T$-dependence in $R_2$ (Fig.~\ref{fig:aster_patch_analysis}c). Therefore, the indirect coupling between inert lipids and ACS via passive lipids leads to the altered transport of the inert lipids, leading to its weak $T-$dependence.

It is worthwhile to make some comments about membrane transport of lipids in light of these observations. First, lipids may move superdiffusively at short times even when they are directly not coupled to active driving forces. Hence, when developing reaction-diffusion-type models for membrane transport, we need to be aware of this fact. Second, as noted earlier~\cite{das2016phase}, lipid segregation is enhanced in the presence of ACS-driven active fluctuations, which may help cells maintain membrane homeostasis in the presence of thermal shock or when thermal fluctuations are not sufficient to create segregated membrane structures, such as lipid rafts.

\section{Discussion}
\par Cells have evolved to deal with their noisy environment. How a cell maintains the robustness of its machinery from the continuous assault of thermal noise has been a matter of extensive investigation \cite{johnson1972thermal, barkai1997robustness, thattai2002attenuation, thattai2004stochastic, bar2006noise}. These studies have repeatedly shown that nonequilibrium driving forces can significantly improve a cell's ability to function. In this paper, we investigate how the transport of molecules, which control and coordinate all cellular processes, are protected against thermal fluctuations. 

We have made two striking observations. First, at the nanometer scale, the flow of the lipids is strongly correlated yet shows clear signs of activated diffusion. On the other hand, the macroscopic activated diffusion of lipids shows no such correlated movements. Where the microscopic correlation is lost is unclear. It is puzzling that even though the lipid movements are strongly correlated at the nanoscale, the standard metric of transport, such as the MSD or the diffusion coefficient, does not show it. A theoretical study would require understanding the many-body transport of lipid clusters that move together at the nanometer scale. The collective motion of the lipid clusters may be random, which is why, perhaps, the MSD or the diffusion coefficient does not capture it. However, we do not have any proof for this hypothesis. We are also unsure about the origin of the weak $T$-dependence in lipid self-diffusion at nanosecond timescales, where such correlated motion is observed. At that scale, it is difficult to calculate the transport coefficients unambiguously, as they are defined in the $\tau\rightarrow \infty$ limit.   

The second striking observation is that nonequilibrium flow from the ACS makes the transport coefficient, $D(T)$ weakly $T$-dependent. Even though we have demonstrated this $T$-independence for a particular nonequilibrium driving, namely the presence of the ACS, it is easy to see that our model is not fine-tuned to any specific biological system. Instead, the key ingredient is the stochastic switching between active and passive moves. As shown in Fig.~\ref{fig:aff_var}, reducing the propensity of the active moves to the passive moves increases $T$-dependence. Therefore, the $T$-independence of $D(T)$ originates from the breaking of detailed balance and is a general observation that is not surprising by itself. Indeed, it is well-known that the fluctuation-dissipation theorem, which relates a transport coefficient to the ambient temperature, is generally violated far from thermal equilibrium \cite{prost2009generalized, horowitz2020thermodynamic, bowick2022symmetry}. What is remarkable is that the $T$-independence of $D(T)$ provides a novel mechanism to ensure the robustness of cell signaling.

Specifically, let's consider the precision of molecular sensing on the cell membrane in the slow diffusion (fast reaction) limit, where the precision of sensing concentration, $c$, of a ligand \cite{bicknell2015limits} varies as
\begin{eqnarray}
    \epsilon (T) = \frac{\delta c}{c} &\propto& \sqrt{\frac{1}{D(T)}}
\end{eqnarray}
Therefore, the precision will decrease exponentially with decreasing temperature in passive membranes, where $D(T)= D_0 e^{-E_A/k_BT}$. For example, consider the precision of ligand detection in a mixed membrane, for which $E_A = 14 k_B \times 310K$ (Fig.~\ref{fig:diffusive_arrhenius}c). We have assumed that the ligand has transport properties similar to those of DPPCs. The precision of detection at 300K will be:
\begin{eqnarray}
    \epsilon(300)/\epsilon(310) &=& \sqrt{D(310)/D(300)}\\
    &=& \sqrt{\exp(-14 + 14*310/300)} \\
    &\approx& 1.26
\end{eqnarray}
That is, there will be a $26\%$ increase in the sensing error. While this is not a dramatic increase in sensing error for an individual receptor, the collective errors are additive and can dramatically alter the precision of signaling. In contrast, $D(T)$ does not change for the active membrane, and the precision remains unchanged over a broader range of temperatures. Hence, we believe $T$-independent transport can provide a potential safeguard against temperature fluctuations. 

$T$-independence in our model arises generically when lipids are transported through correlated flows. Hence, we have reasons to believe that this observation is general. However, it is important to note that our model does not explore all possible variations that can be imagined. For example, we have made the simplifying assumption that all asters are of the same size. In contrast, existing data suggest that asters have a nontrivial size distribution \cite{xia2019nanoscale}. Our data shows that changing the size of the asters, without changing the area fraction, does not change the results qualitatively because it does not change the ratio of the active to passive moves (Fig. S4). In contrast, it is unclear whether the correlated spatial creation and annihilation of the asters can change the results qualitatively. Because it does not change the ratio of the active to passive moves, we have reasons to believe that it should not change the results qualitatively. 

Finally, we have used a new algorithm called barrier hopping dynamics (BHD) to simulate diffusion. We have shown that Kawasaki dynamics, the default choice to model lattice diffusion, does not reproduce the qualitative features of equilibrium diffusion, but BHD does. We have demonstrated that Kawasaki dynamics gives unphysical transport properties. Although BHD reproduces the general features of the equilibrium $D(T)$, the choice of the barriers in this paper is unphysical. That being said, the purpose of introducing BHD was to reproduce the general features of lipid diffusion in thermal equilibrium, which we succeeded in. However, an appropriate choice of diffusion barriers would have allowed us to make quantitative comparisons with the CG-MD model calculations. 

Irrespective of these drawbacks, our calculations provide valuable insights about the transport of lipids on the plasma membrane. It is surprising that even lipids not coupled to the ACS can have $D(T)$ that is almost $T$-independent. As we have shown, a consequence of such temperature-independent transport processes is the robustness of cell sensing. Because of their ubiquity, transport affects not just sensing but all cellular processes. Therefore, it is not hard to imagine that their temperature independence will impact the stability of all cellular processes.

\section{Code Availability}
The in-house scripts for the calculation of MARTINI lipid simulation and lattice simulations are available open-source at \href{https://github.com/argha1992/Transport-on-active-membrane}{github-transport-on-active-membrane}

\section{Acknowledgements}
The Supercomputer Education and Research Center, Indian Institute of Science, Bangalore, India, provided the supercomputer facilities (Param Pravega) to carry out our CG Martini simulations. SGV thanks for the support received through the institute fellowship (010302101122122049) and PMRF fellowship (0202994). AM thanks IISc-IoE fellowship (IE/REAC-22-0112) for funding. SS thanks IISc, Axis Bank Center for Maths and Computing, and SERB-DST (SRG/2022/000163) for funding. The authors would also like to thank Somaditya Santra and Soumyadeep Mondal for their insightful discussions.

\begin{appendix}
    
\section{Methods}
\subsection{CG Molecular Dynamics}\label{CG-MD}
\subsubsection{System} 

For the equilibrium simulation of PM models using CG-MD, we have used three symmetric model membranes with box size $30~nm~\times~30~nm$. The models were built using the INSANE martini tool \cite{wassenaar2015computational} and simulated in GROMACS using MARTINI 3.0 forcefield to achieve faster dynamics \cite{souza2021martini, borges2023martini}. Two membranes were built using pure lipids: (1) the saturated DPPC (di-palmitoyl phosphatidylcholine, $C_{16:0}$) and (2) the polyunsaturated DAPC (di-arachidonoyl phosphatidylcholine, $C_{20:4}$). A third mixed composition membrane (DPPC: DAPC: CHOL, 5:4:3) was also simulated to understand lipid transport in more realistic membranes \cite{lin2019understanding}. All systems were solvated with water, and 0.15 M NaCl was added to attain the physiological ionic condition \cite{wassenaar2015computational}. The system topologies were created in the GROMACS format. Five different temperatures (298, 302, 306, 310, and 314 K) were used to check $T$ dependency of the PM models. All the production simulation (Table. \ref{tab:Simulation Details}) was conducted for 5 $\mu$s, except two pure DPPC systems that were run for 30 $\mu$s using GROMACS-2022.2 \cite{van2005gromacs, bekker1993gromacs, abraham2015gromacs} with 20 fs timestep where each 100 ps snapshots were stored for analysis. The detailed simulation protocol is given in the SI. 

\begin{table}[h!]
    \centering
    \begin{tabular}{ccccc}
        \hline
        PM & Replicas & Time & Time  & Total Time \\
         & & ($\mu$s/run) & ($\mu$s/temp) & ($\mu$s) \\
        \hline
        Pure DPPC & 2 & 30 & 60 & 300 \\
                  & 6 & 5 & 30 & 150 \\
        Pure DAPC & 10 & 5 & 50 & 250 \\
        Mixed     & 12 & 5 & 60 & 300 \\
        \hline
        Total     &    &   &    & 1000 \\
    \end{tabular}
    \caption{Simulation Details}
    \label{tab:Simulation Details}
\end{table}

\subsubsection{Lipid flow simulation} 
Another set of simulations was conducted to study the nanometer scale correlated flows of lipids. The $T$ range was kept the same (298-314 K), and only the pure DPPC model was simulated. The starting structures were taken from the long simulations (after 3 $\mu$s) to their respective temperatures. All the parameters were kept the same, but the run time was reduced to 1 ns with 20 fs timesteps. These short-time simulations were run for 20 replicas at each temperature, and each 20 fs coordinates were stored. 

\subsubsection{Analysis} 
The analysis was conducted with in-house developed Python 3.11 programs using the MDAnalysis package \cite{gowers2016mdanalysis} and GROMACS-2022.2 tools \cite{abraham2015gromacs} and visualization was performed with the molecular graphics viewer VMD \cite{humphrey1996vmd}. 

\subsection{MSD calculation}\label{MSD_cal}

\par To understand the $T$-dependence of the lipid transport, we have employed various analysis protocols to measure the diffusivities of lipids. 

\subsubsection{Bulk diffusivity measurement}
\par The MSD for lipids at different temperatures was calculated over the last 2 $\mu$s of each 5 $\mu$s runs, whereas the last 5 $\mu$s were used for the 30 $\mu$s trajectories. The analyzed part of the trajectories was further divided into 1 $\mu$s blocks \cite{ingolfsson2014lipid, venable2017lipid} to calculate MSD. The MSD is defined as:
\begin{equation}\label{eq:3}
\langle X^2 \rangle = \langle | \mathbf{r}_i^\tau - \mathbf{r}_i^0 |^2 \rangle = \frac{1}{N} \sum_{i=1}^{N} [ (x_i^\tau - x_i^0)^2 + (y_i^\tau - y_i^0)^2 ],
\end{equation}
where $\langle X^2 \rangle$ is the MSD, N is the number of particles in the system, $x_i^\tau,~y_i^\tau$ and $x_i^0,~y_i^0$ is the position of the $i^{th}$ particle on XY plane at times $\tau$ and 0, respectively. 
\par The {\verb |gmx msd|} tool was used to calculate the MSD of the $PO_{4}$ (PC lipids) and ROH (for CHOL) beads of the lipids (only the headgroup of lipids were considered for the MSD calculation). The results were plotted on a log-log scale to determine the power-law dependence of diffusion. The diffusion or transport coefficient was calculated by multiplying the slope with the coarse-grained (CG) conversion factor of 4, as suggested in \cite{marrink2004coarse, marrink2007martini}. Subsequently, the transport coefficients at different temperatures were plotted to derive the activation energy of specific lipid molecules using the Arrhenius equation:
\begin{equation}\label{eq:4}
	D(T) = D_0 e^{-\frac{E_A}{k_BT}},
\end{equation}
\begin{equation}\label{eq:5}
	\ln D(T) = -{\frac{1}{T}}{\frac{E_A}{k_B}} + \ln D_0,
\end{equation}

where $D_0$ is a $T$-independent constant that depends on the lipid properties, $k_B$ is the Boltzmann constant, and $E_A$ is the activation energy. $E_A$ was determined by calculating the slope of $\ln D(T)$ vs $1/T$ curve for different PM models. Per experimental convention, $1000/T$ was used instead of $1/T$ for the plots \cite{bag2014temperature}. The errors were estimated from the standard deviation among the replicas.

\subsubsection{Local diffusivity measurement}
\par To understand the short-time spatial behavior of lipids, we analyzed one ns simulation trajectories. The system was divided into grid boxes measuring $3~nm \times~3~nm$, each containing approximately 10-15 $PO_4$ beads of DPPC lipids, which were tracked to measure the local MSD (Fig. S3). The lag time $\tau$ was varied between 0 and 1 ns with a resolution of 20 fs. The time-averaged MSD for the $PO_4$ groups within each grid box was computed, which showed subdiffusive transport with the anomalous exponent of 0.6. Therefore, the transport coefficient was not the diffusion constant. However, we could still determine the transport coefficient, $D(T)$, from the slope of the MSD curve (check the results section for details). Following the $D(T)$ calculation, we analyzed its distribution across the grids for the abovementioned temperatures. This method was repeated for the remaining replicas, and we compared the spatial average of $D(T)$ at different $T$.

\subsection{Correlation analysis}\label{correlation_cal}
\par To accurately quantify the correlation in the lipid displacements (see results), we measured the displacement-displacement correlation (DDC) function for DPPC \cite{donati1999growing, narumi2009study, puosi2012spatial}. DDC generally reveals the degree of dynamical heterogeneity in the system, indicating the spatial distribution of its mobile elements \cite{ediger2000spatially, richert2002heterogeneous}. DDC is used extensively to quantify dynamical heterogeneity in the glass transition \cite{poole1998spatial, bennemann1999growing, narumi2008simulation}.

\par On a fixed time interval $\tau$, the displacement vectors for the $i^{th}$ and the $j^{th}$ lipids, $\hat{{x}}_i({r}_i,\tau)$ and $\hat{{x}}_j({r}_j,\tau)$ were calculated. The correlation function is then given by:
\begin{equation}
        {g(r, \tau)_x = \frac{1}{N} \sum_{i=1}^{N} \sum_{j=1}^{N} \langle \hat{{x}}_i({r}_i,\tau) \cdot \hat{{x}}_j({r}_j,\tau)\delta(r-|r_{ij}(t_0)|) \rangle},
\end{equation}
where N is the number of particles in the system, $r_{ij}(t_0)$ is the initial separation distance between $i^{th}$ and the $j^{th}$ lipids.

\subsection{Lattice model simulation}\label{lattice model}

In the lattice model \cite{machta2011minimal,kimchi2018ion,das2016phase}, every grid is occupied by a lipid species that either couple to the underlying ACS or does not couple. We assume that the lipid that couples to the ACS does not influence its behavior, and it gets \textit{passively} transported by the ACS. Following the terminology of ref. \cite{gowrishankar2012active}, we call them \textit{passive} lipids. The lipids that do not interact with the ACS are called \textit{inert} lipids. The lattice model is a coarse-grained model, and each grid is averaged over many molecular lipids. Therefore, the passive-inert classification has to be judged based on the net interaction of the lipids in a grid. Our lattice model can, in principle \cite{kang1989dynamic, fichthorn1991theoretical}, be derived from a systematic coarse-graining of a detailed molecular model, which is beyond the scope of the current paper. Here, we assume that such a coarse-graining is possible, and the effective interaction between the lattice-lipids is of similar strength as that between two molecular lipids, which are of the order of $0.5~k_BT$. In doing so, we lose all the chemical details of the CG model, which constrains us from providing any legitimate macroscopic thermodynamic information about the ACS-lipid interaction. However, the lattice model can provide a qualitative understanding of the transport process in the presence of ACS, even when detailed molecular information is missing. Unless otherwise stated, from now on, we use the term lipids to imply the lattice lipids when talking about the lattice model.

\subsubsection{Interaction between lipids}
 Lipids of the same type (passive-passive or inert-inert) interact with each other through attractive Lennard-Jones (LJ) potential $U(r)= -J(\frac{\sigma}{r})^{6}$ with a cutoff radius of 2.5$\sigma$, where we took one lattice unit as $2^{1/6} \sigma$ for excluded volume interaction between unlike lipids. Hence, we have at most two lattice units ($=2 \times 2^{1/6}\sigma < 2.5 \sigma$) attractive interaction between like lipids as shown in Fig.~\ref{fig:KD_vs_BHD}c. The interaction strength $J$ is chosen so that $J/k_BT = 1/1.43 $ at $T = 293~K$. Lipids of different species interact with excluded volume interaction only, such that $J = 0$ for unlike lipids. Hence, the total interaction energy, $U$, of a given pair of lipids varies between $-6\mbox{ to } 0~k_BT$, which is in the same ballpark as the activation energies calculated from the molecular simulation. The neighbor configurations and the interaction energy distribution ($-U$) for different possible neighbor configurations are shown in Fig. \ref{fig:KD_vs_BHD}d.

\subsubsection{Equilibration}
 To equilibrate the lipids on the lattice, we used Kawasaki exchange moves \cite{landau2021guide}, which exchange two neighboring lipids with a probability $p$ determined by the following formula: 
\begin{eqnarray}
	p = \begin{cases}
		1 & \mbox{if } \Delta E \leq 0 \\
		e^{-\Delta E/k_BT} & \mbox{if } \Delta E > 0,
	\end{cases}
\end{eqnarray} 
where $\Delta E$ is the difference in the interaction energy before and after the swap. Kawasaki moves guarantee the system reaches the correct equilibrium state if runs sufficiently long. However, due to kinetic effects, \textit{local} Kawasaki moves often lead to an arrested state. Hence, we used \textit{global} Kawasaki moves for equilibrating a random initial lipid configuration. In the global algorithm, two lipids were chosen randomly from the entire system and swapped using the same criteria, which avoided the kinetic traps and rapidly converged to the global equilibrium state at a given temperature. 

\subsubsection{Diffusive (passive) moves} \label{BHD_moves}
The equilibrated state served as the initial condition for the simulations to study the transport of the lipids. 
 
The algorithm, which we call \textit{barrier hopping dynamics} or BHD, is as follows: 
\begin{enumerate}
	\item Pick a random lipid and a neighboring lipid
	\item Calculate the total interaction energy of the lipid and the neighboring lipid, $U$. 
	\item Measure the barrier height $B = -U$. 
	\item Swap the lipids with probability  $ p = e^{-B/k_BT}$. 
\end{enumerate}

Barrier heights chosen this way are unrealistic as they do not capture the various entropic effects that determine the free energy of the transition state. Obtaining the ``correct" barrier heights would require detailed molecular simulation followed by systematic coarse-graining of the barrier heights, which is beyond the scope of this manuscript~\cite{kang1989dynamic, fichthorn1991theoretical}. Instead, our model provides a simple alternative that can be used to understand the effect of active fluctuations on lipid transport, which is the main focus of this paper. Indeed, similar models have been used previously with similar intentions \cite{landau2021guide, uebing1991monte, uebing1994determination}. Another concern is that if $U$ is positive, such as in the presence of long-range repulsive forces, $B$ will become negative. There, we need to identify a better definition for $B$. Here, we do not consider long-range repulsive forces, and $B$ is always non-negative.  

\subsection{Detailed-balance breaking active moves}\label{active_moves}
Because of the interaction of the motor proteins with the ACS, various out-of-equilibrium structures form spontaneously. One of the most prominent structures is the asters, radially symmetric concentrations of actin filaments, typically $\sim 100~nm$ in size. It has been shown that the asters are central hubs of PM-ACS interaction \cite{gowrishankar2012active}. Therefore, in the lattice model, we consider the effect of asters on the lipid transport on the plasma membrane. Specifically, we assume the asters are disk-like dynamic regions on the membrane, where the lipids can be transported via active moves. The details of the aster dynamics are outlined later. 

Once a passive lipid is inside the circular region defined by the aster, it can be advected by active moves if it also binds to the ACS. Because passive lipids stochastically bind to the ACS, we assume that the passive lipids have some binding probability, $p_{b}$, which controls how often they bind to the ACS. A passive lipid bound to the ACS is advected to its core \cite{das2016phase}. The detailed algorithm and simulation parameters are given in the SI.

\subsubsection{Aster Remodeling}
Simultaneously with lipid dynamics, asters follow life-death processes with remodeling rate($\tau_a$) such that the area fraction of asters ($A_{frac}$) remains constant. The number of aster discs at any time is the same and defined by $N_{aster} = \frac{A_{frac} L^2}{\pi R_a^2}$. Once asters are initialized, a lifetime ($t_{life}$) sampled from an exponential distribution with mean lifetime ($\tau_a$) \cite{sarkar2023lifetime} is designated to each aster from the time of birth ($t_{birth}$). No overlap between asters was ensured while carrying out the aster birth-death process. When $t-t_{birth}=t_{life}$ aster disappears and appears with some other independent location and lifetime.

\subsection{Algorithm for active moves}

The detailed balance-breaking or active moves are implemented on the lattice through Kinetic Monte Carlo exchange moves, slightly modified from \cite{das2016phase}. The algorithm chosen to perform the exchange move is as follows:

\begin{enumerate}
    \item A lipid is chosen randomly from $L\times L$ sites.
    
    \item Next, a neighbor is chosen for the exchange. 
    
    \begin{itemize}
        \item If the chosen lipid is passive and is bound to the ACS (with binding probability $p_{b}$), then the neighbor lipid for exchange is selected to lie in the aster core direction with probability $p > 0.25$. Other neighbors are chosen with probability $(1-p)/3$.

        \item Otherwise, any of the four neighboring lipids are chosen with equal probability. 
    \end{itemize}
    
    \item $p_b$ is defined from the two-state model of the (un)bound states of passive lipid with ACS. Hence, $p_b = 1/(1+e^{\frac{\Delta E}{k_BT}})$, where $\Delta E$ is the energy difference between the bound and the unbound state. 
    
    \item Let $\vec{r_i}$, $\vec{r_f}$ and $\vec{r_a}$ be the position vectors of the chosen passive lipid, neighbor site for exchange, and the aster core, respectively. To perform the directional move on the lattice, we have defined three cases from $\vec{r_a}-\vec{r_i}=\Delta\vec{ r}= \Delta x \hat{x}+ \Delta y \hat{y}$, where $\Delta\vec{ r}$ is the position of aster core w.r.t the chosen passive lipid.
    \begin{enumerate}[label=(\alph*)]
        \item $|\Delta x| > |\Delta y| $: The position of exchange site is given by, $\vec{r_f}=$ ($x_i + sign(\Delta x), y_i$).
        \item $|\Delta x| < |\Delta y| $: Similarly, $\vec{r_f}=$ ($x_i,y_i + sign(\Delta y)$).
        \item $|\Delta x| = |\Delta y| $: The direct diagonal move is not possible. Therefore,  ($x_i + sign(\Delta x), y_i$) or ($x_i,y_i + sign(\Delta y)$) is chosen randomly.
    \end{enumerate}

    \item Also, passive lipids near the outer aster edge feel interaction with ACS, and radial move is performed with probability $(1-p_X)p_{b}$ where X is the species lying on the aster core before the swap. $p_X = 0$ if $X$ is an inert lipid, else it is $p_b$. 
    
    \item If a passive lipid is found on the aster core, it cannot move on its own, and no exchange move was performed.
    
    \item For passive lipids away from any aster, BHD move is performed.
\end{enumerate}
\end{appendix}

\newpage
\clearpage
\setcounter{equation}{0}
\setcounter{figure}{0}
\setcounter{table}{0}
\setcounter{page}{1}
\makeatletter
\renewcommand{\theequation}{S\arabic{equation}}
\renewcommand{\thetable}{S\Roman{table}}
\renewcommand{\thefigure}{S\arabic{figure}}
\setcounter{section}{0}
\begin{center}
\textbf{\large Supplementary Information: Self-diffusion is temperature independent on active membranes}\\
\text{\large Saurav G. Varma, Argha Mitra, Sumantra Sarkar}\\
\textit{Center for Condensed Matter Theory, Department of Physics, Indian Institute of Science, Bengaluru, Karnataka, 560012}\\
\end{center}

\section{CG-MD simulation protocol}
\paragraph{Energy minimization and equilibration:} The system's energy was minimized using the steepest descent and the conjugate gradients methods in the next step. We ran at least 5000 steps of energy minimization to ensure proper geometry and to minimize steric clashes. Following energy minimization, all systems were subjected to a six-step equilibration protocol (Table:~\ref{tab:Equilibration Details}), where the stiffness of the positional restraint on the lipid head groups ($PO_4$ bead of DPPC and DAPC) was in gradually decreased to avoid collapse of the system. Five different temperatures (298, 302, 306, 310, and 314 K) were controlled by velocity-rescale thermostats \cite{bussi2007canonical} with a coupling constant of 0.1 ps, which was used. For the equilibration, we used Berendsen barostat \cite{berendsen1984molecular} with the compressibility of $4.5~\times10^{-5}~bar^{-1}$ semi-isotropic scaling \cite{marrink2007martini, marrink2013perspective}.
\begin{table}[h!]
    \centering
    \begin{tabular}{cccc}
        \hline
        Step & Restraint ($KJmol^{-1}s^{-1}$) & Timestep (fs) & Time (ns) \\
        \hline
        1 & 200 & 2 & 1 \\
        2 & 100 & 5 & 1 \\
        3 & 50 & 10 & 1 \\
        4 & 20 & 15 & 0.75 \\
        5 & 10 & 20 & 1 \\
        6 & 2 & 20 & 50 \\
        \hline
        Total & & & 54.75 \\
    \end{tabular}
    \caption{Equilibration Details}
    \label{tab:Equilibration Details}
\end{table}
  
\paragraph{Production runs:} For these simulations velocity rescale temperature bath was applied \cite{bussi2007canonical}. The pressure was controlled with the Parrinello-Rahman barostat \cite{parrinello1981polymorphic} with a 12 ps coupling constant, and the compressibility of $3~\times10^{-4}~bar^{-1}$ \cite{de2016martini}. However, due to the longer spatial scale during production, the chance of undulation increases, which was significantly reduced by using a mild positional restraint ($2~kJmol^{-1}nm^{-1}$) on the $PO_4$ beads in the Z-direction~\cite{ingolfsson2014lipid, ingolfsson2017computational}. All the production simulation was conducted for 5 $\mu$s, except two pure DPPC systems that were run for 30 $\mu$s using GROMACS-2022.2 \cite{van2005gromacs, bekker1993gromacs, abraham2015gromacs} with 20 fs timestep where each 100 ps snapshots were stored for analysis. The cumulative simulation time was 1 ms.

\newpage
\section{Lattice model simulation parameters}
\begin{table}[h!]
    \centering
    \begin{tabular}{cc}
        \hline
        Simulation parameters & Values \\ 
        \hline
        Lattice size ($L \times L$) & $100\times 100$\\
        Radius of aster ($R_a$) & 8\\
        Areal density of asters ($A_{frac}$) & 0.2 \\
        Mean lifetime of aster remodeling ($\tau_a$) & 10\\
        Equilibration steps& 500 steps \\
        Time to reach steady state & 1000 steps\\
        Production run & 18500 steps \\
        No. of replicas for each ($T$) & 40 \\
        Probability to move towards aster core (passive lipids) ($p$) & 0.8 \\ 
        Temperatures ($T$) & 281 - 321K (spacing = 3K)\\
        Binding probability of ACS to PM ($p_b$) & 0.0 - 1.0 (spacing = 0.2) \\
        Interaction strength of LJ ($J/k_BT$ at T=293 K) & 1/1.43 (like lipids) \& 0 (unlike lipids)\\
        LJ interaction cutoff radius & $2.5 \sigma$ \\
        1 lattice unit & $2^{1/6} \sigma$ \\
        \hline
    \end{tabular}
    \caption{The mesoscopic lattice model simulation is performed on a 2D square lattice of size $L \times L$ with periodic boundary conditions. The lattice sites are occupied with either passive or inert lipids in equal ratios. All spatial parameters are in the units of lattice constant and time scales are in the units of MC steps. 1 lattice unit ($\sigma_d$)= 2.5/100 $\mu$m \cite{das2016phase} and 1 MC step ($\tau_d$) = $\frac{\sigma_d^2}{4D}$ = 0.446 ms, where D $\sim$ 14 $\mu m^2 s^{-1}$ is the typical diffusivity value from our CG-MD simulations.}
    \label{tab:neq_details}
\end{table}

\section{Movie Details}
Movie: "displace\_with\_activity.mp4" shows the active dynamics of lipids on a lattice at 311 K. The passive lipid densities are shown in red and the inert lipid densities are shown in blue.  Passive tracer lipid tracks are shown in maroon and inert tracer lipid tracks are in blue. Asters are marked by black circles.

\newpage
\clearpage 
\begin{figure}[ht]
    \centering
    \includegraphics[width=\columnwidth]{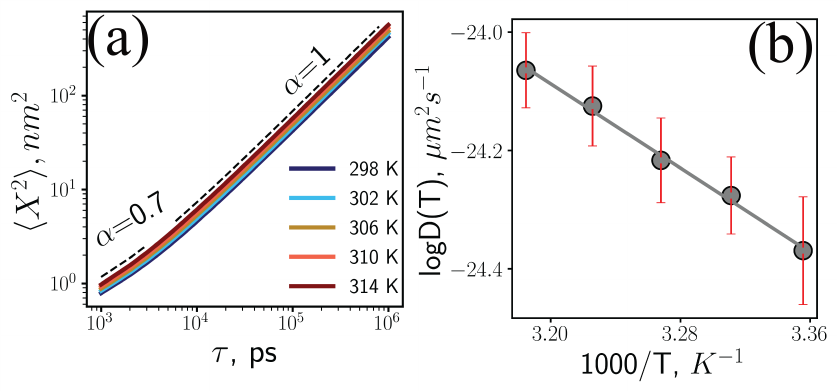}
    \caption{In equilibrium, DAPC lipids show diffusive behavior, and the diffusion coefficient follows Arrhenius kinetics. (a-b) MSD ($\langle X^2 \rangle$) and diffusivity (semilog scale) of pure DAPC membrane at different $T$ (298 K - 314 K). All the results from long-timescale simulations. Error bars are the standard deviation of 10 replicates.}
    \label{fig:diffusive_arrhenius_dapc}
\end{figure}

\begin{figure}[ht]
    \centering
    \includegraphics[width=0.75\columnwidth]{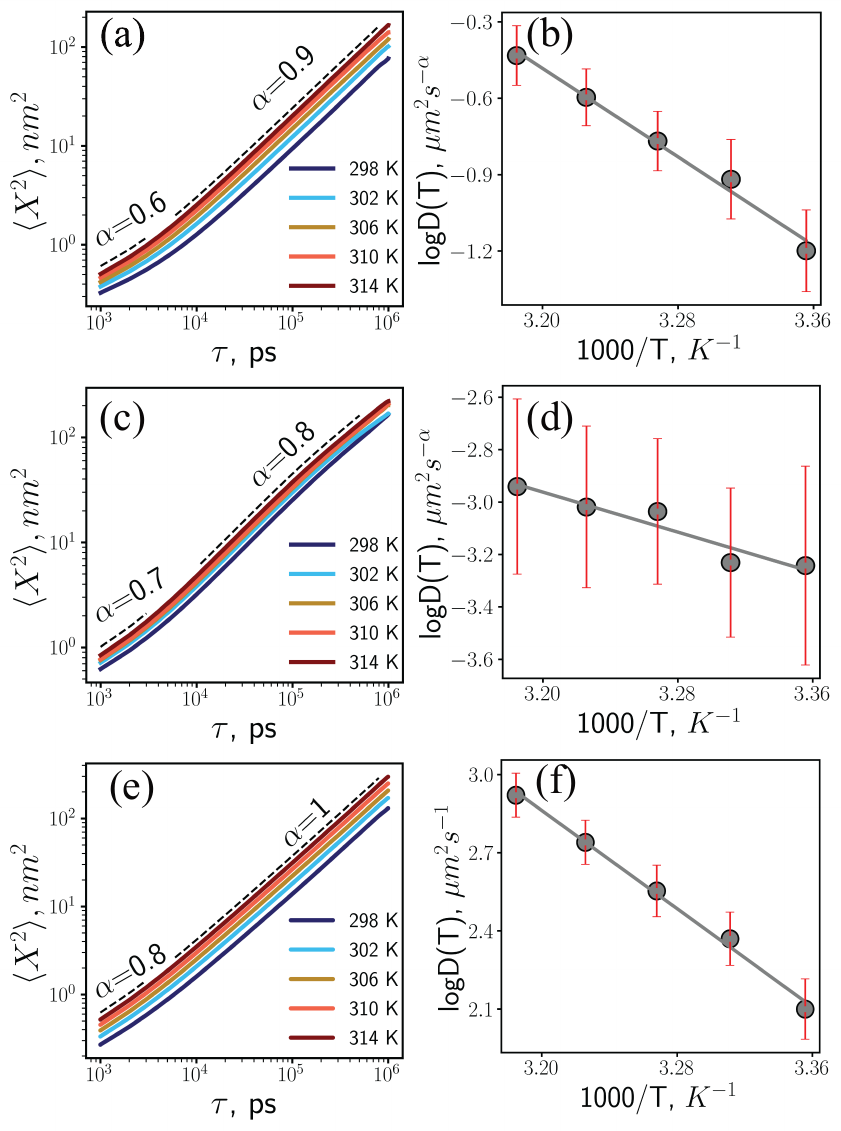}
    \caption{In equilibrium, mixed model PC lipids show subdiffusive behavior (For DPPC, $\alpha$ = 0.9 and DAPC, $\alpha$ = 0.8) and CHOL shows diffusive behavior. MSD ($\langle X^2 \rangle$) and transport coefficient (semilog scale) of DPPC (a-b), DAPC (c-d), and for CHOL (e-f) in the mixed model membrane at different $T$ (298 K - 314 K). All the results from long-timescale simulations. Error bars are the standard deviation of 12 replicates.}
    \label{fig:diffusive_arrhenius_mixed_model}
\end{figure}

\begin{figure}[ht]
    \centering
    \includegraphics[width=\columnwidth]{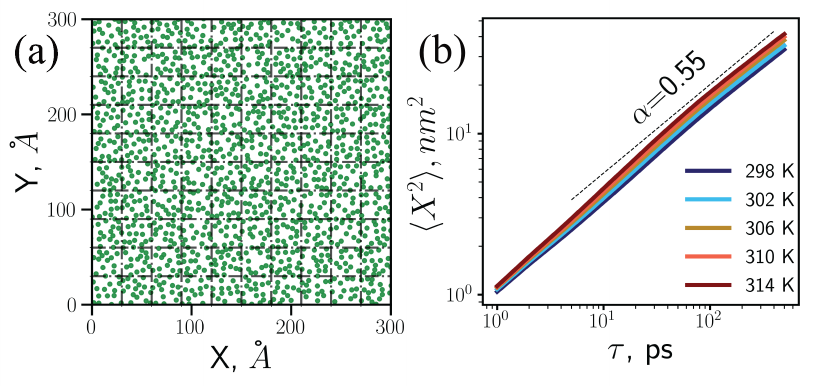}
    \caption{Nanoscale simulation system snapshot and MSD.  (a) Snapshot $PO_4$ groups of DPPC at 314 K (each grid = $3 \times 3$ nm). (b) MSD ($\langle X^2 \rangle$) averaged over grids at different $T$ (298 K - 314 K).}
    \label{fig:nanoscale_MSD}
\end{figure}

\begin{figure}[ht]
    \centering
    \includegraphics[width=\columnwidth]{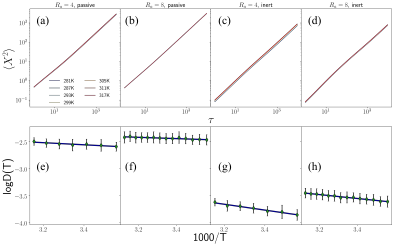}
    \caption{Change in aster radius has no effect on diffusive behavior of lipids. The data from two different sizes of asters ($R_a =4$ and 8) is compared in (a)-(h) for passive and inert lipids. }
    \label{fig:aster_radius_compare}
\end{figure}

\newpage
\clearpage
\bibliography{references}
\end{document}